\documentclass[preprint,12pt,authoryear]{elsarticle}

\usepackage{amssymb}
\usepackage{hyperref}

\journal{New Astronomy Reviews}

\begin{document}

\begin{frontmatter}

\title{The \textit{Gaia} white dwarf revolution}

\author[inst1]{Pier-Emmanuel Tremblay}

\affiliation[inst1]{organization={Department of Physics, University of Warwick},
            city={Coventry},
            postcode={CV4 7AL}, 
            country={UK}}
            
\author[inst1]{Antoine B\'edard}
\author[inst1]{Mairi W. O'Brien}
\author[inst1]{James Munday}
\author[inst1]{Abbigail K. Elms}
\author[inst2]{Nicola Pietro Gentillo Fusillo}
\author[inst1]{Snehalata Sahu}

\affiliation[inst2]{organization={Universita degi studi di Trieste},
            addressline={Via Valerio, 2}, 
            city={Trieste},
            postcode={34127}, 
            country={Italy}}

\begin{abstract}
This review highlights the role of the \textit{Gaia} space mission in transforming white dwarf research. These stellar remnants constitute 5-7\% of the local stellar population in volume, yet before \textit{Gaia} the lack of trigonometric parallaxes hindered their identification. The mission's Data Release 2 in 2018 provided the first unbiased colour-absolute magnitude diagram of the local stellar population, identifying 260\,000 white dwarfs, with the number later increasing to over 355\,000 in Data Release 3. Since then, more than 400 white dwarf studies have made critical use of \textit{Gaia} data, establishing it as a fundamental resource for white dwarf identification, fundamental parameter determination and more recently spectral type characterisation. The review underscores the routine reliance on \textit{Gaia} parallaxes and extensive use of its photometry in white dwarf surveys. We also discuss recent discoveries firmly grounded in \textit{Gaia} data, including white dwarf mergers, exotic compact binaries and evolved planetary systems. 
\end{abstract}




\end{frontmatter}


\section{Introduction}
\label{sec:intro}

At the end of the life cycle of a star less massive than 8--10\,$M_{\odot}$, the remaining core, composed mostly of carbon, oxygen, and in some cases neon, contracts into a white dwarf. These are the most common stellar remnants, accounting for 5-7\% of the local stellar population in volume \citep{Gaia-nearby-stars}. White dwarfs are characterized by their Earth-like radii and correspondingly low luminosities. In the local volume, their luminosity distribution peaks within the range of $10^{-3}$ to $5 \times 10^{-5}$\,$L_{\odot}$. They gradually cool over time, and their cooling track in a Hertzsprung-Russell (HR) diagram has long been used to date stellar populations.

Until \textit{Gaia}, local stellar volume samples were limited to $\approx$500 white dwarfs within $\approx$50\,pc of the Sun \citep{Limoges2015,Holberg2016}, and several thousands more at larger distances identified from Sloan Digital Sky Survey (SDSS) spectroscopy \citep{York2000,Kepler2019}. This situation arose from the challenge of distinguishing cool white dwarfs from main-sequence stars based solely on photometry, without the aid of a distance estimate. The sample of white dwarfs with trigonometric parallax measurements was itself limited to about $\approx$200 objects \citep{Bergeron2001,Subasavage2017,Bedard2017,Leggett2018}, with some of the most precise constraints originating from the 20 white dwarfs observed by \textit{Hipparcos} \citep{Vauclair1997}. Decades of effort have enabled the definition and comparison of two independent methods for acquiring white dwarf parameters (effective temperature -- $T_{\rm eff}$, mass, radius, luminosity, surface gravity -- $\log g$, surface chemical composition and cooling age): the photometric and astrometric technique \citep{Koester1979,Bergeron2001}; and the spectroscopic method \citep{Bergeron1992}. The large independence of these two approaches is anchored in the well-constrained white dwarf mass-radius relation \citep{Parsons2017,Tremblay2017,Bedard2017}.

Since the launch of \textit{Gaia} in 2013, the mission has been collecting a comprehensive all-sky sample of astrometric data (position, parallax, proper motion) and apparent magnitudes in three broad optical filters ($G$, $G_{\rm BP}$, $G_{\rm RP}$; \citealt{Gaia2016}). Its Data Release 1 (DR1) sample included only six directly observed white dwarfs but several more in wide binaries for which the companion had a parallax \citep{Tremblay2017}.
The situation changed dramatically on 25 April 2018, when \textit{Gaia} DR2 was able to draw for the first time the HR diagram of the local stellar population including over 260\,000 white dwarfs \citep{Gaia2018,Jimenez2018,Gentile2019}. This number later increased to over 355\,000 white dwarfs in \textit{Gaia} EDR3/DR3\footnote{DR3 largely contributed to adding auxiliary data, such as low-resolution spectrophotomery, to white dwarfs already identified in EDR3. We refer to this as the DR3 white dwarf sample.}. Since then, more than 400 white dwarf studies have used \textit{Gaia} data\footnote{Refereed papers from 2018 onward including both ``Gaia" and ``white dwarfs" in their abstract according to SAO/NASA Astrophysics Data System.} as a fundamental input. The focus of this review is to illustrate the progress made in white dwarf research thanks to \textit{Gaia} DR2 and DR3 data through these recent studies. For further background, we refer to the bibliography of this work, as well as several reviews on white dwarf evolution, atmospheres, magnetic fields, constitutive physics and pulsations \citep{Fontaine2001,Althaus2010,Ferrario2020,Saumon2022,Bedard2024}, as well as their evolved planetary systems \citep{Farihi2016,Veras2021,Xu2024}.

It is now routine for white dwarf studies to rely on \textit{Gaia} parallaxes as it is by far the largest and most precise such dataset. \textit{Gaia} (spectro-)photometry also features prominently in recent white dwarf characterisations. This review emphasises the significance of survey papers utilising extensive \textit{Gaia} datasets but also noteworthy are high-profile findings on rare individual white dwarfs, such as mergers \citep{Caiazzo2023}, exotic compact binaries \citep{Burdge2022} and evolved planetary systems \citep{Vanderburg2020}, all firmly grounded in \textit{Gaia} data. 
We acknowledge the significant contributions of the \textit{Gaia} DPAC team, whose crucial work and papers have allowed the community results presented herein.

\section{\textit{Gaia} white dwarf samples}
\label{sec:sample}
\begin{figure}
    \centering
    	\includegraphics[width=1.0\textwidth]{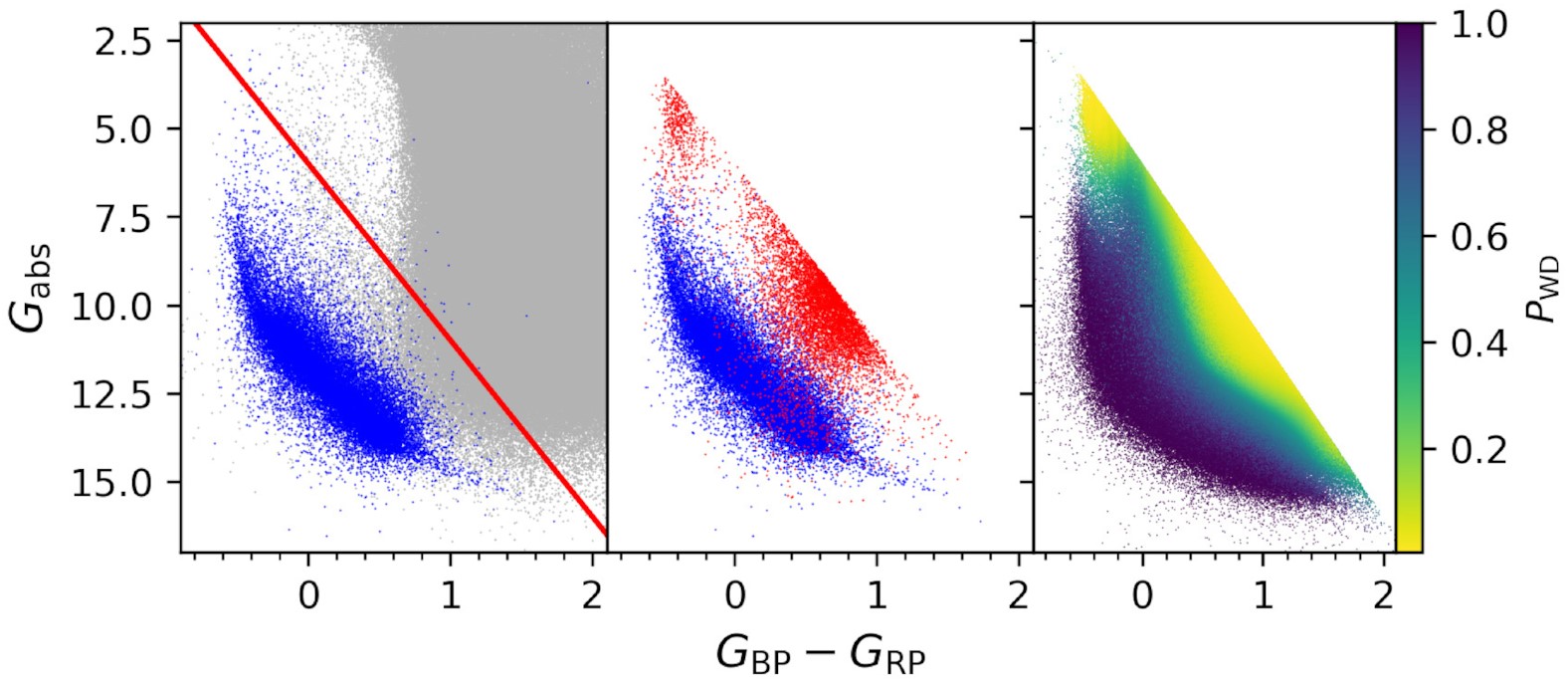}
	\caption{Example of \textit{Gaia} EDR3 white dwarf selection from \citet{Gentile2021}. \textit{Left}: HR diagram showing a representative sample of 2 million random sources with a parallax precision better than 100\% (gray points). 
The blue points represent spectroscopically confirmed white dwarfs from SDSS defining the white dwarf locus. A broad cut (red line) is used to limit the number of contaminants. \textit{Center}: Distribution of
SDSS white dwarfs (blue) and contaminants (red) included in the final selected region of the HR diagram. \textit{Right}: Catalogue of 1\,280\,266 white dwarf candidates, with the colour scale illustrating the probability of being a white dwarf ($P_{\rm WD}$) (Source: \citealt{Gentile2021}).}
    \label{fig:Selection}
\end{figure}

\textit{Gaia} data enable the characterisation of white dwarf optical colors (e.g., $G_{\rm BP} - G_{\rm RP}$ in magnitude units), which are primarily determined by the $T_{\rm eff}$ and atmospheric composition. Additionally, \textit{Gaia} photometry and parallax provide absolute magnitudes (e.g., $G_{\rm abs}$), which depend on the $T_{\rm eff}$, atmospheric composition and radius. Both colours and magnitudes also have a small dependence on $\log g$, interior composition and interstellar extinction. Using the $T_{\rm eff}$-dependent white dwarf mass-radius relation, the position of a white dwarf in the \textit{Gaia} HR diagram can be translated into a mass and luminosity. White dwarfs are characterised by faint absolute magnitudes for a given colour, hence are found in a region of the HR diagram that is predicted to be otherwise empty, at least in principle for sufficiently high data quality (Fig.\,\ref{fig:Selection}). We remind the reader that white dwarfs with larger masses have smaller radii, so those with increasingly high masses are located in the bottom left corner of the HR diagram.

Starting with \textit{Gaia} DR2, different teams including \textit{Gaia} DPAC have extracted \textit{Gaia} white dwarf samples \citep{Gaia2018,Jimenez2018,Gentile2019,Pelisoli2019}, culminating in the current DR3 samples \citep{Gaia-nearby-stars,Gentile2021,Jimenez2023}. In this section we highlight the most recent white dwarf samples, as the selection process was qualitatively the same for DR2 and DR3.

Despite the apparently isolated white dwarf locus in the HR diagram, several difficulties arose in the selection of white dwarf samples for two primary reasons: statistical and systematic \textit{Gaia} data uncertainties, as well as binary stars. Fig.\,\ref{fig:Selection} confirms that a simple astrometric precision cut of \textsc{parallax\_over\_error} $>$ 1 is not sufficient to differentiate between white dwarfs and main sequence stars. However, the intrinsic faintness of white dwarfs can actually offer an advantage in selecting them from \textit{Gaia}, as absolute magnitudes of $M_{\rm G}\approx12$ and 13.5\,mag, which are the median for magnitude and volume-limited samples, respectively, imply parallaxes of 2.5--5\,mas at $G=20$\,mag. This is higher than the \textit{Gaia} average parallax precision of $\approx$0.5\,mas at the limiting magnitude of $G\approx20$\,mag \citep{Gaia2023}, hence a cutoff in parallax significance (\textsc{parallax\_over\_error}) has been used as a compromise to retain a majority of \textit{Gaia} white dwarfs while eliminating most other sources. Additionally, it has been shown that white dwarf candidates can also be selected through \textit{Gaia} precise reduced proper-motion method, even if parallax significance is below 1$\sigma$ \citep{Gentile2021}. 

Systematic issues in \textit{Gaia} data have also impacted white dwarf selection, namely blue color excesses (through the parameter \textsc{phot\_bp\_rp\_excess\_factor}) especially in crowded regions like the Galatic plane and the Magellanic Clouds, resulting in red objects such as nearby M dwarfs and brown dwarfs contaminating the white dwarf sample. There are also issues with spurious large parallax values and high astrometric noise (through parameters \textsc{astrometric\_sigma5d\_max}, \textsc{astrometric\_excess\_noise} or \textsc{ruwe}). Different methods have been proposed, both by the \textit{Gaia} DPAC team and white dwarf community, to filter these non-white dwarf sources (see e.g. \citealt{Gentile2021}).

\begin{figure*}
    \centering
    	\includegraphics[width=0.6\textwidth]{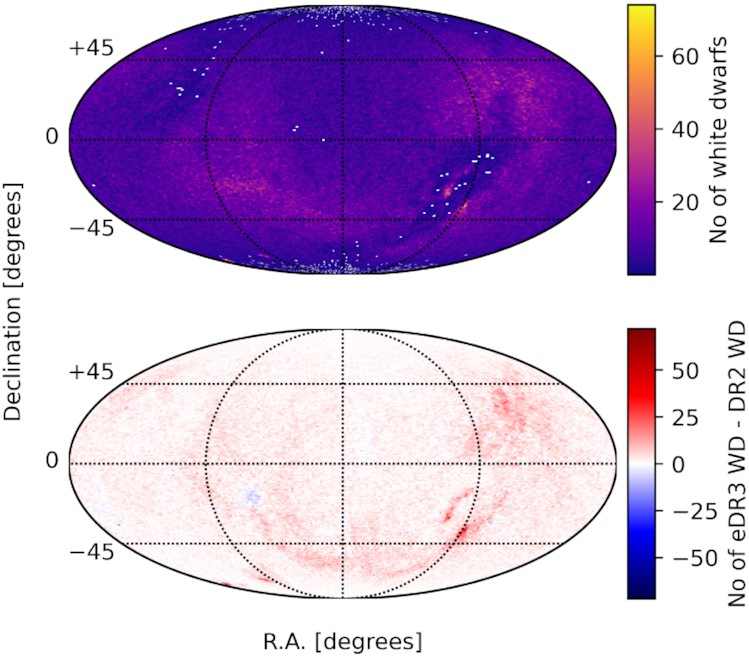}
	\caption{\textit{Top}: Sky density (in 1 deg$^{2}$ bins) of high probability \textit{Gaia} EDR3 white dwarf candidates ($P_{\rm WD} >$ 0.75). \textit{Bottom}: Difference between EDR3 and DR2 (Source: \citealt{Gentile2021}).}
    \label{fig:Sky}
\end{figure*}

By attempting to select a maximum number of \textit{Gaia} white dwarfs, it is inevitable that they will overlap with other sources in the HR diagram. Hence, different approaches have been used to define probabilities of a source being a white dwarf, either through a direct comparison of its uncertainty contour in HR diagram compared to the locus of known white dwarfs \citep[][see Fig.\,\ref{fig:Selection}]{Gentile2019,Gentile2021}, or through machine learning \citep{Gaia-nearby-stars}. We note that while these \textit{Gaia} white dwarf catalogues allow the definition of large white dwarf samples with a small contamination level, spectroscopy is needed to confirm the nature of individual white dwarf candidates.

White dwarf selection in the HR diagram is also complicated by unresolved binaries covering the full space between the white dwarf locus and the main sequence. Considerable effort has been invested in identifying binary systems including a bright, extremely low-mass white dwarf candidate \citep{Pelisoli2019,Pelisoli2019_sdA}, as well as unresolved white dwarf + main-sequence (hereafter WD+MS) binaries \citep{Pala2020,Rebassa2021}. Compared to single white dwarfs, these systems are pushed towards the main-sequence branch of the \textit{Gaia} HR diagram and have a lower space density, resulting in the fraction of contaminants being much larger, and the need for more stringent quality or volume cuts. Hot white dwarfs ($T_{\rm eff} > 30\,000$\,K) are also very close to the hot subdwarfs in the HR diagram \citep{Culpan2022}, resulting in a partial overlap of these candidates even at relatively high parallax precision.

Community work has allowed the identification of $\approx$359\,000 high probability \textit{Gaia} white dwarf candidates over all sky (Fig.\,\ref{fig:Sky}). It is estimated that this represents 67--93\% of all white dwarfs with $G<20$\,mag \citep{Gentile2021}. Completeness is worse for hot, distant white dwarfs and better for cool, nearby sources. In particular, it is estimated that \textit{Gaia} white dwarf completeness within 40\,pc is $>$97\% \citep{Hollands2018_Gaia,O'Brien2024}.

\subsection{Volume-limited samples}
\label{sec:volume}

The high \textit{Gaia} source completeness enables the creation of representative white dwarf samples through the use of selection functions sampling the full cooling track in the HR diagram \citep{Rix2021}. However, local volume-complete samples centred around 20--100\,pc of the Sun offer several advantages: they are based on decades of archival spectroscopic and photometric observations; have a much higher rate of confirmed versus candidate white dwarfs; and are the only volumes where faint and old white dwarfs (cooling age $\gtrsim$ 6\,Gyr) can be observed. The white dwarf with the faintest known luminosity, WD\,J2147$-$4035, which has one of the most evolved planetary systems around a white dwarf known to date \citep{Elms2022}, has an absolute magnitude of $M_{\rm G}=17.73$\,mag and a distance of 27.9\,pc - implying it would not be detected by \textit{Gaia} at distances larger than $\approx$40\,pc (apparent $G>20.7$\,mag). Local volumes of white dwarfs are essential for studies of stellar evolution, old planetary systems and Galactic star formation history.

\begin{figure}
    \centering
    	\includegraphics[width=0.8\textwidth]{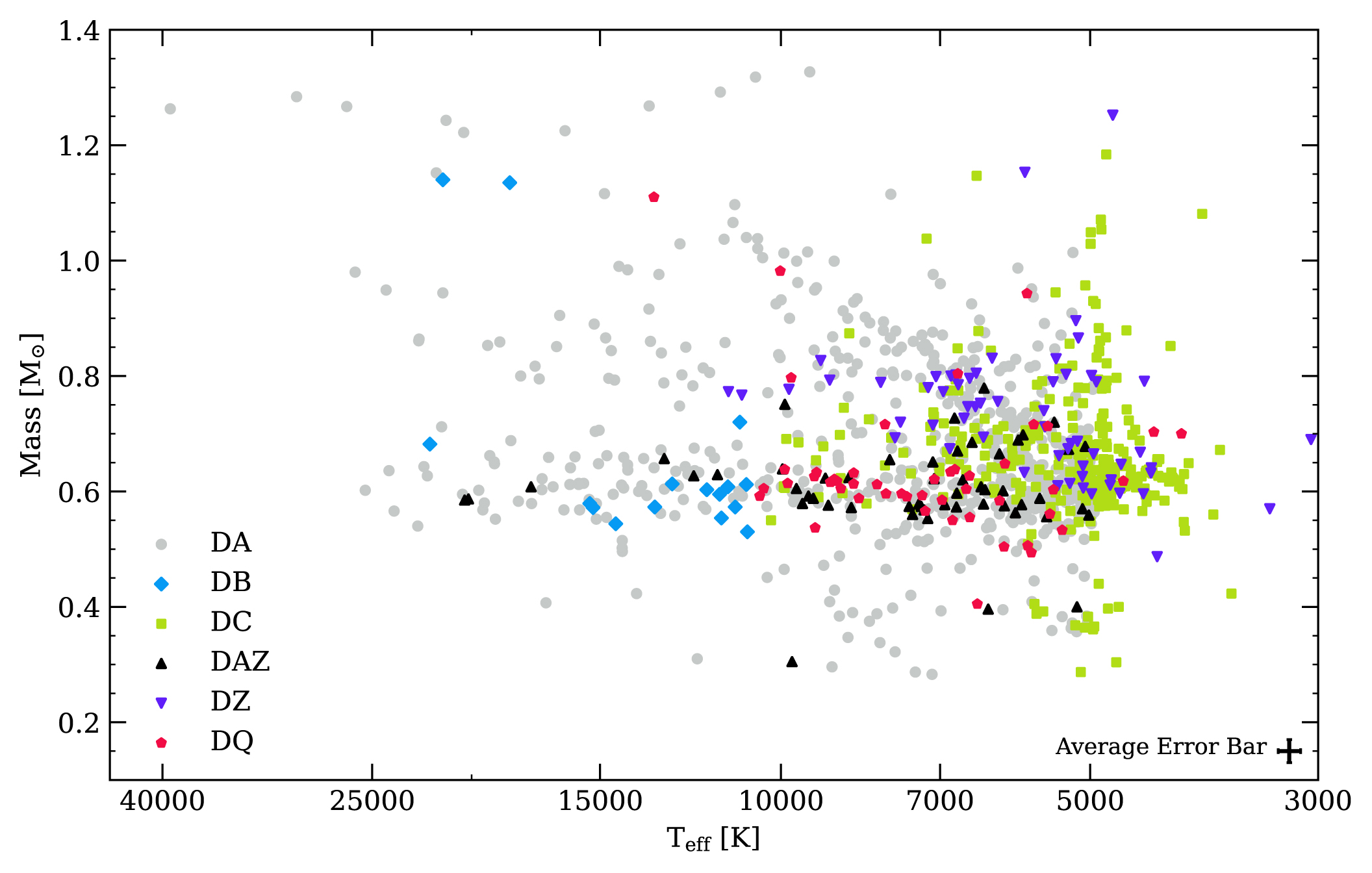}
	\caption{Model-atmosphere derived mass and $T_{\rm eff}$ for all 40\,pc white dwarfs in DR3 as determined from \textit{Gaia} photometry and astrometry. A correction to the parameters is implemented below 6000\,K to account for model atmosphere systematics. The spectral types are indicated by the shape and colour of points, with ``D" for degenerate, ``A" for hydrogen Balmer lines, ``B" for helium lines, ``C" for no lines, ``Z" for metals except carbon, and ``Q" for carbon.  More complex spectral types are simplified to their most prominent features. The average statistical error is shown on the lower right of the panel (Source: \citealt{O'Brien2024}).}
    \label{fig:40pc}
\end{figure}

The nearly complete 40\,pc volume of spectroscopically confirmed white dwarfs was recently defined thanks to both pre-\textit{Gaia} work (e.g. \citealt{Giammichele2012}; \citealt{Limoges2015}) and dedicated follow-up observations (\citealt{Hollands2018_Gaia}; \citealt{Tremblay2020}; \citealt{Mccleery2020}; \citealt{OBrien2023}). The \textit{Gaia} DR3 defined 40\,pc sample of \citet{O'Brien2024} contains 1076 spectroscopically confirmed white dwarfs out of 1081 candidates from \citet{Gentile2021}, hence has a completeness of 99 per cent in white dwarf spectral types at medium-resolution (Fig.\,\ref{fig:40pc}). However, there is a minimum of $\approx$40 additional white dwarfs that are found within this volume but are not identified as white dwarf candidates by \citet{Gentile2021}, the majority of which are in unresolved WD+MS binaries \citep{O'Brien2024,Golovin2023}.

Larger volume samples are being populated by multi-object medium-resolution spectroscopic surveys \citep{SDSSV,4MOST,DESI2022,WEAVE} and \textit{Gaia} DR3 low-resolution spectrophotometry, hence more white dwarf candidates are now being confirmed. 
In particular, the \textit{Gaia}-SDSS-I/IV footprint within 100\,pc is nearly complete for $T_{\rm eff} > 7000$\,K \citep{Kilic2020_100pc}.

\section{White dwarfs in the HR diagram}
\label{sec:HR}
\begin{figure}
    \centering
    	\includegraphics[width=0.49\textwidth]{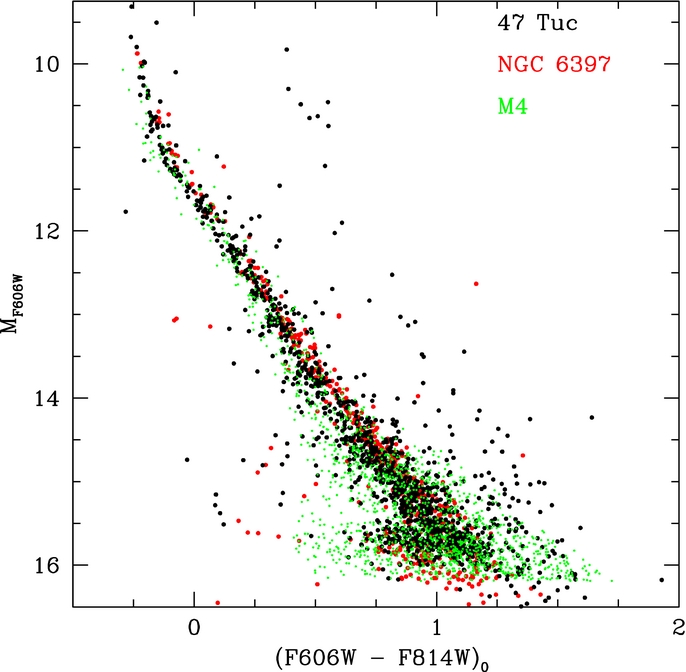}
	\includegraphics[width=0.49\textwidth]{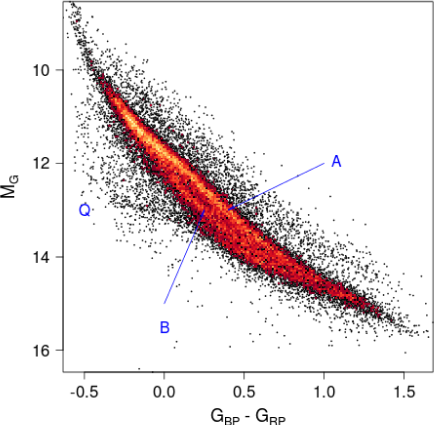}
	\caption{\textit{Left}: Overlay of the white dwarf cooling in the globular clusters 47 Tuc, NGC 6397, and M4, as observed by the Hubble Space Telescope (Source: \citealt{Richer2013}). \textit{Right}: \textit{Gaia} DR2 HR diagram of 26\,264 white dwarfs with a parallax precision better than 5\%. The A-, B- and Q-branches are discussed in the text (Source: \citealt{Gaia2018}).}
    \label{fig:HR}
\end{figure}

Predictions of white dwarf cooling tracks in the HR diagram \citep{VanHorn1968,Fontaine2001,Althaus2010,Bedard2020} have played a pivotal role in establishing connections between white dwarf populations and their Galactic environment. Before the advent of \textit{Gaia}, determining the position of white dwarfs on the HR diagram relied on indirect methods involving model-dependent distance estimates, except for observations of globular clusters which contained a sufficient amount of white dwarfs that are all located at essentially the same distance. As such, \textit{Gaia} DR2 \citep{Gaia2018} introduced the inaugural HR diagram for field white dwarfs, revealing substantial differences from the globular cluster population (refer to Fig.\,\ref{fig:HR}). Most notably, the \textit{Gaia} HR diagram revealed the mysterious Q- and B-branches, at first unexplained, and not previously seen in the globular cluster populations \citep{Richer2013}.

The fundamental differences between white dwarfs observed in globular clusters and \textit{Gaia} HR diagrams are age distributions and completeness. In globular clusters, most white dwarfs visible on the cooling sequence have formed from long-lived Sun-like stars and have a mass of $\approx$0.53\,M$_{\odot}$. White dwarfs formed from more massive and shorter-lived main-sequence stars all have large cooling ages, and are too faint to be seen or are only seen at the very bottom of the HR diagram \citep{Richer2013}. In contrast, the \textit{Gaia} HR diagram contains stars formed at a nearly constant rate over the last $\approx$10\,Gyr, resulting in white dwarfs of all possible ages and masses populating the diagram. We now review in turn the properties of the \textit{Gaia} Q- and B-branches.

\subsection{Crystallisation (Q-Branch)}
\label{sec:crystallisation}

The Q-branch illustrated in Fig.\,\ref{fig:HR} has been interpreted as arising from the interior crystallisation of white dwarfs and associated physical processes \citep{Tremblay2019b}. It had long been predicted that the carbon, oxygen and neon core of a white dwarf, as it cools, attains a critical temperature leading to a phase transition from liquid to solid \citep{VanHorn1968}. This first-order phase transition releases latent heat as well as gravitational energy due to chemical separation, resulting in a cooling delay of $\approx$1 Gyr \citep{Althaus2012,Blouin2020,Bauer2023}, potentially observable in the HR diagram as a bottleneck or overdensity of white dwarfs \citep{VanHorn1968}. Given that core crystallisation occurs at higher temperatures for higher masses, it is anticipated to generate a branch not aligned with the cooling tracks but, instead, aligned with the observed Q-branch.  However, these well-known energy sources are insufficient to explain the observed overdensity, indicating that white dwarfs experience an additional $\sim$1 Gyr cooling delay as they crystallise \citep{Tremblay2019b,Bergeron2019,Blouin2020,Kilic2020_100pc}.  White dwarfs in wide binaries, with both components formed at the same time, present an intriguing observational avenue for further constraining crystallisation delays if one component has started the process \citep{Heintz2022,Venner2023}.

\begin{figure}
    \centering
    	\includegraphics[width=0.49\textwidth]{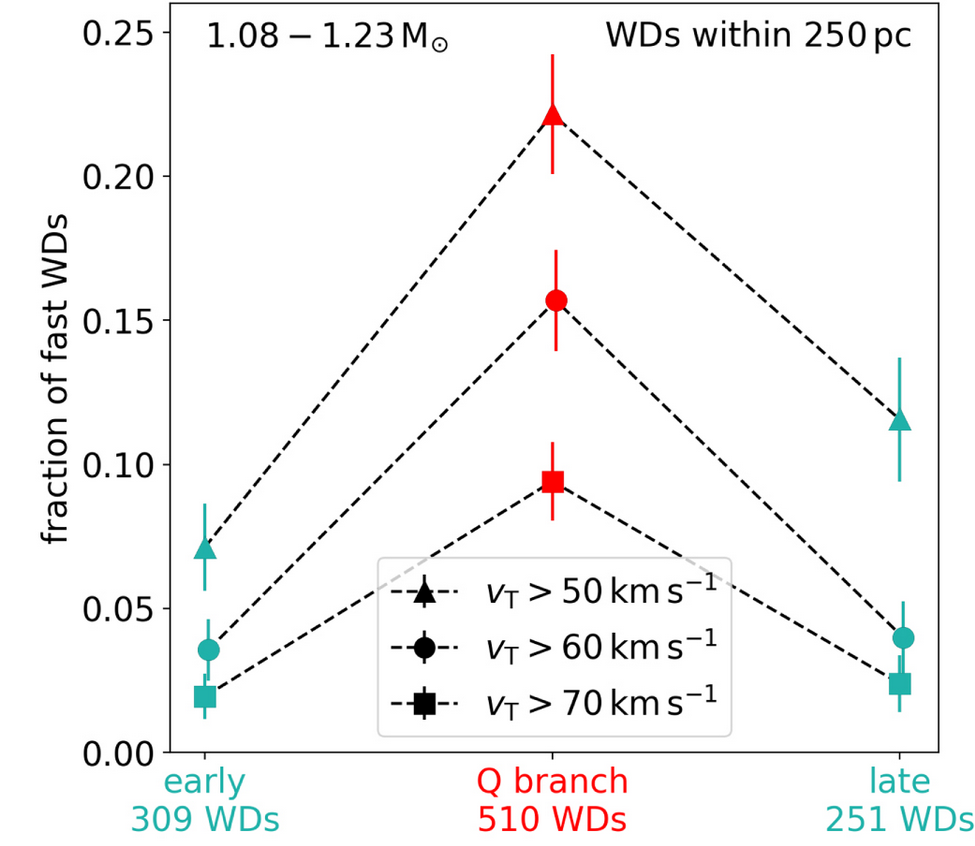}
	\caption{The fraction of fast moving white dwarfs for the mass range 1.08-1.23\,$M_{\odot}$ for different tangential velocity cuts. There are significantly more fast-moving white dwarfs on the Q-branch than both above and below in the HR diagram. Fast white dwarfs are interpreted to be older than slow moving white dwarfs according to the Galactic disc age–velocity-dispersion
relation. The high fraction of fast white dwarfs on the Q-branch has been interpreted as an extra cooling delay (Source: \citealt{Cheng2019}).}
    \label{fig:kinematic}
\end{figure}

Further insight into this problem was provided by the kinematic distribution of ultra-massive white dwarfs (1.08--1.23\,$M_{\odot}$) at the blue end of the Q-branch \citep[][see Fig.\,\ref{fig:kinematic}]{Cheng2019}. The authors demonstrated that in this mass range, \textit{Gaia} data do not support a short cooling delay for all white dwarfs, but instead a prolonged cooling delay of $\gtrsim$8\,Gyr for 5--9\% of white dwarfs. This discovery has given rise to a substantial theoretical literature attempting to identify the nature of the powerful energy source that essentially halts the cooling \citep{Bauer2020,Caplan2020,Horowitz2020,Blouin2021,Blouin2021b,Caplan2021,Camisassa2021}. In most cases, the extra-delayed objects have been interpreted as merger products with a distinct internal chemical composition allowing for the prolonged delay, particularly a carbon/oxygen-dominated core with a high fraction of neutron-rich impurities such as $^{22}$Ne. The most promising solution involves enrichment in $^{22}$Ne and $^{26}$Mg by mergers of white dwarfs and subgiant stars \citep{Shen2023}, and the subsequent distillation of these elements during the crystallisation process \citep{Blouin2021}. In essence, the crystals are predicted to be depleted in neutron-rich species and thus lighter than the surrounding liquid, implying that they float up and melt while heavier liquid is gradually displaced downward, thereby releasing a large amount of gravitational energy. White dwarf evolution models including this distillation mechanism indeed predict that the cooling is halted for $\sim$10 Gyr and hence successfully explain the extra-delayed population of ultra-massive white dwarfs \citep{Bedard2024_nature}. Besides, we note that a variant of the distillation scenario for standard core compositions could also potentially explain the shorter cooling delay experienced by lower-mass white dwarfs originating from either mergers or single star evolution \citep{Blouin2021}.

\subsection{Spectral evolution (B-Branch)}
\label{sec:spectral}

The A- and B-branches delineated in the \textit{Gaia} HR diagram were promptly recognised as distinct trajectories representing the cooling of hydrogen-rich and helium-rich atmosphere white dwarfs within the temperature range of 7000--11\,000\,K \citep{Gaia2018, Gentile2019}. The prevalent A-branch consists in white dwarfs of DA spectral type characterised by their optical Balmer lines. The notably divergent B-branch is made of helium-rich atmosphere white dwarfs, predominantly of DC, DZ, and DQ spectral types (indicating no lines, metal lines, or carbon lines, respectively; see also Fig.\,\ref{fig:40pc}). However, the large deviation between A- and B-branches cannot be explained by predicted cooling tracks based on pure-helium and pure-hydrogen model atmospheres.

The first explanation was proposed by \citet{Bergeron2019}, who demonstrated that trace hydrogen with an abundance in number of H/He = $10^{-5}$ in helium-rich atmospheres could account for the bifurcation. Such hydrogen contamination is too small to be observed in most currently available spectroscopic observations, although the explanation remains unsatisfactory since both warmer and cooler helium-rich atmospheres are generally inconsistent with an abundance of H/He = $10^{-5}$ \citep{Bergeron2022, Elms2022, Blouin2023}.

\begin{figure}
    \centering
    	\includegraphics[width=0.9\textwidth]{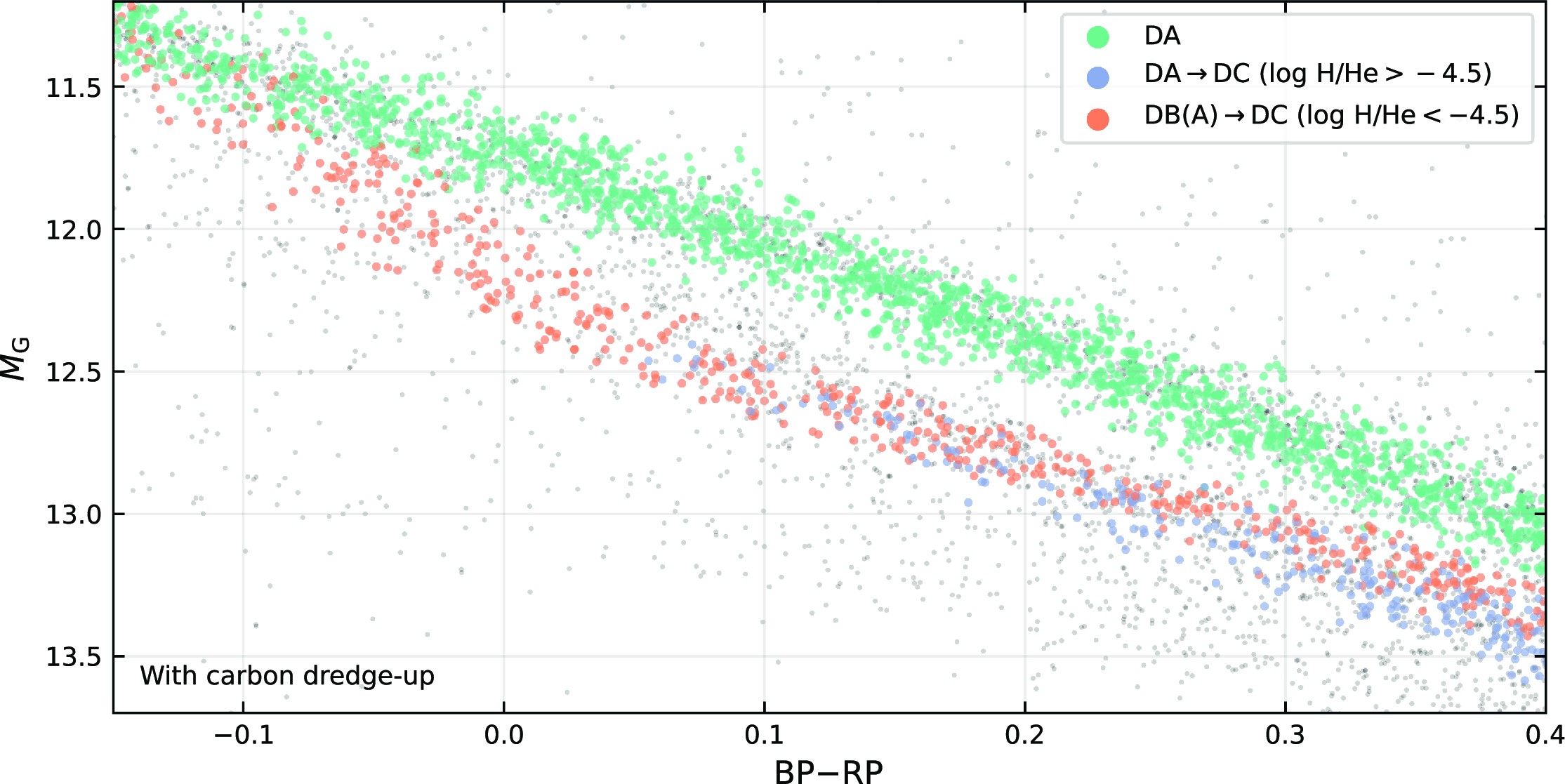}
	\caption{Gaia HR diagram of white dwarfs in the 100\,pc sample (small grey dots). The coloured points show the simulation of a population of cooling pure-hydrogen (green) and helium-rich (with carbon dredge-up) white dwarf atmospheres that can explain the B-branch bifurcation. The blue and orange points for the helium-rich population distinguish objects that have and have not experienced convective mixing, respectively (Source: \citealt{Blouin2023}).}
    \label{fig:B-branch}
\end{figure}

Meanwhile, investigations on the \textit{Gaia} population of helium-atmosphere DQ white dwarfs with trace carbon \citep{Koester2019,Coutu2019,Blouin2019c,Koester2020,Kawka2023}\footnote{These studies have highlighted a class of massive and warm DQ white dwarfs, possibly linked to stellar mergers and discussed in Section\,\ref{sec:binaries}, and a class of average-mass cool DQ white dwarfs discussed in this section.} have led to the hypothesis that cool DQ white dwarfs may represent the upper part of the carbon pollution range resulting from convective dredge-up \citep{Bedard2022b}. In that scenario, most helium-rich DC white dwarfs may also harbour spectroscopically undetected lower traces of carbon. This scenario was recently tested through white dwarf population synthesis \citep{Camisassa2023,Blouin2023} and appears to provide a more plausible explanation for the B-branch bifurcation than trace hydrogen, as initially suggested by \citet{Bergeron2019}. Both effects are largely degenerate for individual optical white dwarf observations, as both trace carbon and hydrogen contribute to increasing the number of free electrons and, consequently, He$^{-}$ opacity. However, trace carbon offers a better fit to UV photometric observations \citep{Blouin2023b} and to the narrowness of the B-branch in the HR diagram \citep[][see Fig.\,\ref{fig:B-branch}]{Blouin2023}.

Various studies have investigated the closely related question of white dwarf spectral evolution from \textit{Gaia} samples, examining changes in atmospheric composition during cooling. These studies largely confirmed pre-\textit{Gaia} results, indicating that white dwarfs are born with diverse atmospheric compositions \citep{Werner2014,Bedard2020,Reindl2023} and then evolve into a maximum hydrogen-atmosphere DA white dwarf fraction of 90--95\% within the so-called DB gap from $\approx$40\,000 to $\approx$20\,000\,K \citep{Genest2019b,Ourique2019,Bedard2020,Torres2023,Vincent2023}, a phenomenon attributed to element diffusion \citep{Althaus2005,Bedard2022,Bedard2023}. \textit{Gaia} has also confirmed and better quantified the increase in the incidence of helium-atmosphere white dwarfs below 30\,000 K \citep{Lopez2022,Torres2023,Vincent2023,O'Brien2024}. The phenomenon is most likely explained by convective dilution and mixing \citep{Rolland2018,Rolland2020,Cunningham2020,Bedard2022,Bedard2023}. The fraction of helium-atmosphere white dwarfs seems to plateau at $\approx$30\% for $T_{\rm eff} < 9000$\,K, with no major spectral evolution events beyond this point \citep{Blouin2019b,Elms2022,O'Brien2024}, although this is a topic of ongoing debate \citep{Caron2023}. Spectral evolution due to the accretion of planetary debris is discussed in Section\,\ref{sec:pollution}. A more extensive review of white dwarf spectral evolution is presented in \citet{Bedard2024}.

We conclude this section with the reminder that the Q- and B-branches are not observed in the HR diagram consisting of white dwarf populations in globular clusters (left panel of Fig.\,\ref{fig:HR}). The explanation in the former case is relatively straightforward since lower mass white dwarfs found in globular clusters have not yet solidified. The absence of the B-branch, however, remains a mystery and may suggest a not-yet understood metallicity, mass or environmental dependence in the production of helium-rich white dwarf atmospheres. 

\section{White dwarf parameters}
\label{sec:parameters}

\textit{Gaia} has brought about a revolution in determining white dwarf parameters. 
The interpretation of the \textit{Gaia} HR diagram, as discussed in the previous section, did rely on white dwarf atmosphere and evolution models to predict colours and absolute magnitudes. In addition, several studies have explicitly derived individual \textit{Gaia} white dwarf parameters by fitting photometry and astrometry, typically utilising synthetic fluxes from model atmospheres and the mass-radius relation specific to white dwarfs (see e.g. \citealt{Gentile2021}). In principle, this framework could enable the derivation of all fundamental white dwarf parameters solely from \textit{Gaia} data. However, practical challenges arise due to the necessity of defining atmospheric chemical composition and interstellar extinction. Currently, these input parameters are largely constrained by external spectroscopy and extinction maps, respectively.

Recent studies have derived the spectral types of over 100\,000 white dwarfs from \textit{Gaia} DR3 spectrophotometry \citep{Spectrophoto2023,Jimenez2023,Garcia-Zamora2023,Torres2023,Vincent2023}, paving the way for more self-consistent parameter determinations using \textit{Gaia} data alone. Nevertheless, for a significant portion of white dwarfs exhibiting weak lines, magnetic fields, or carbon and metal pollution, accurate parameter determinations will continue to require external medium- or high-resolution spectroscopic data (\citealt{Blouin2019};\citealt{Coutu2019}; \citealt{Koester2019}; \citealt{Hardy2023}; \citealt{Caron2023}; \citealt{O'Brien2024}).

The \textit{Gaia} dataset encompasses the largest number of white dwarfs, spanning all conceivable spectral types, and is distinctive for possessing the only substantial sample of parallaxes. Consequently, \textit{Gaia} data serve as a benchmark for testing white dwarf parameter determinations employing different methods, as discussed herein. Many of these studies have relied on well-behaved hydrogen- and helium-line white dwarfs to facilitate comparisons.

{\bf \textit{Gaia} versus optical photometry:} Comparisons with ground-based and narrower band SDSS, Pan-STARRS, J-PLUS and SkyMapper photometry have been extensively performed \citep{Gentile2019,Bergeron2019,Lopez2019, Mccleery2020,Lopez2022,Izquierdo2023,Sahu2023}. Most of these comparisons rely on the same \textit{Gaia} astrometric data and mass-radius relation, leaving $T_{\rm eff}$ as the sole independent parameter. In most studies, a good agreement at the 1--2 percent level in $T_{\rm eff}$ has been observed for cool white dwarfs in the range 6000--10\,000\,K \citep[see e.g. figure 3 of][for Pan-STARRS]{Mccleery2020}. We note that optical colours become less sensitive to $T_{\rm eff}$ as the latter increases, with the scatter between \textit{Gaia} and Pan-STARRS reaching a minimum of 10\% at $T_{\rm eff} \approx 40\,000$\,K \citep{Gentile2019}. For cool white dwarfs of all spectral types below $\approx$6000\,K, \textit{Gaia} astrometry has led to the identification of a significant low-mass problem, with photometric masses of white dwarfs at $\approx$4000\,K being too small by $20\%$ compared to predictions from Galactic population models \citep{Hollands2018_Gaia,Blouin2019b,Bergeron2019,Mccleery2020,Caron2023,Cukanovaite2023,O'Brien2024,Cunningham2024}. Inaccuracies of microscopic opacities in these dense, cool white dwarf atmospheres have been identified as a likely culprit \citep{Saumon2022,Caron2023,O'Brien2024}.

\begin{figure}
    \centering
    	\includegraphics[width=0.49\textwidth]{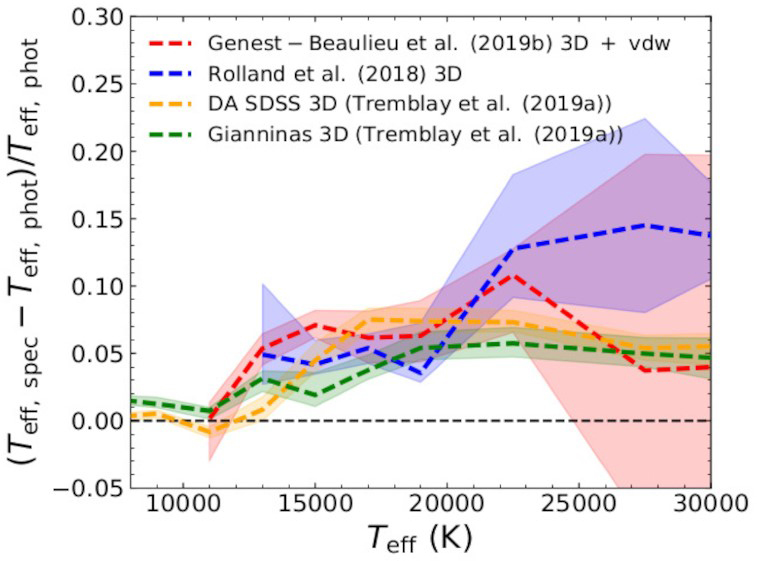}
    	\includegraphics[width=0.49\textwidth]{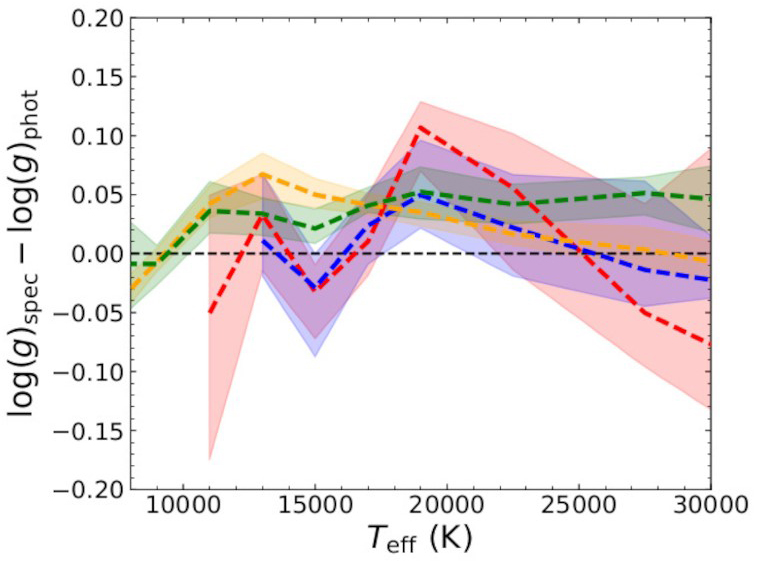}
	\caption{A comparison between \textit{Gaia} DR3 derived $T_{\rm eff}$ and $\log g$ values and spectroscopically derived parameters from the literature. The DB(A) spectroscopic parameters are from \citet{Genest2019b,Rolland2018} with 3D convection corrections \citep{Cukanovaite2021}. The DA spectroscopic parameters are from \citet{Gianninas2011,Tremblay2019} with 3D convection corrections \citep{Tremblay2013}. The dashed lines indicate the median offset and the coloured areas represent the error on the median (Source: \citealt{Gentile2020}).}
    \label{fig:spectro}
\end{figure}

{\bf \textit{Gaia} versus optical spectroscopy (line fitting):} Extensive work has been performed to compare \textit{Gaia} photometric parameters to independent spectroscopic fits of hydrogen Balmer or helium lines \citep{Tremblay2019,Genest-Beaulieu2019,Narayan2019,Gentile2020,Tremblay2020,Gentile2021,Cukanovaite2021,OBrien2023,Izquierdo2023,Axelrod2023}. These studies have consistently revealed a systematic offset between the spectroscopic and photometric determinations of $T_{\rm eff}$ and $\log g$ for both DA and DB(A) white dwarfs. The spectroscopic solutions yield temperatures that are 2-5\% higher and surface gravities that are 0.02-0.05 dex greater (refer to Fig.\,\ref{fig:spectro}). It has been suggested that inaccuracies in the physics of line broadening could be a contributing factor to the observed offset since most spectroscopic predictions are based on the same broadening theories. However, Fig.\,\ref{fig:spectro} shows that the offset is similar for DA and DB white dwarfs, exhibiting entirely different microphysics and atomic lines in their spectra. If the similarity in the offset between these two types of white dwarfs is not coincidental, this could indicate that the issue lies in the calibration of photometric measurements.

{\bf \textit{Gaia} versus other wavelengths:} Medium-resolution far-UV spectrophometry conducted with the Cosmic Origins Spectrograph (COS) aboard the Hubble Space Telescope (\textit{HST}) has also been compared to \textit{Gaia} parameters \citep{Sahu2023}. Intriguingly, the findings indicate that COS temperatures are lower than both optical spectroscopic temperatures (by 3\%) and \textit{Gaia} temperatures (by 1.5\%). The analysis of mass distributions further reveals that COS masses are smaller by approximately 0.05 and 0.02\,$M_{\odot}$ compared to Balmer lines and photometric masses, respectively. Recent studies have also shown the potential of DA white dwarfs as infrared flux calibrators \citep{Gentile2020}. 

{\bf Hybrid photometric and spectroscopic methods:} Studies have added systematic errors, based on the offsets noted above, to simultaneously fit photometry, astrometry and spectroscopy \citep[see e.g.][]{Hollands2023}. 
The \textit{HST} CALSPEC flux scale (\citealt{Bohlin2014}; \citealt{Maiz2018}; \citealt{Bohlin2019}) has also been used as part of a recipe to re-calibrate photometry for self-consistent hybrid fits, therefore enhancing the precision of white dwarf parameters \citep{Narayan2019,Gentile2020,Axelrod2023}. 

{\bf Other methods:} The comparison between spectroscopic gravitational redshifts and \textit{Gaia} masses and radii has been explored in recent research \citep{Arseneau2023}. An innovative white dwarf mass measurement technique for \textit{Gaia} has been introduced by \citet{Hwang2023}, demonstrating the feasibility of determining dynamical masses through wide binaries. Moreover, \textit{Gaia} has facilitated the creation of new white dwarf catalogues, enabling the search for pulsating white dwarfs in time-domain surveys \citep{Vincent2020,Guidry2021,Romero2022,Steen2024}. This, in turn, has led to improved determinations of asteroseismic masses and temperatures, offering valuable comparisons with \textit{Gaia} parameters \citep{Romero2022}.

\subsection{White dwarfs in binaries}
\label{sec:binaries}

{\bf Resolved (wide) binaries:} The precise astrometry from \textit{Gaia} has significantly expanded the scope of samples for wide binaries. \citet{El-Badry2021} have compiled a catalogue featuring $\approx$1400 WD+WD and $\approx$16\,000 WD+MS wide binaries, which has been important in comparative analyses with binary population synthesis models. This revealed a notable deficit of WD+WD binaries in comparison to predictions \citep[see also][]{Toonen2017,O'Brien2024}. \citet{El-Badry2018} found a break in the separation distribution at 3000\,au and 1500\,au for WD+MS and WD+WD pairs, respectively, in significant contrast with the flatter MS+MS distribution. The authors attributed this result to white dwarfs experiencing a kick of approximately 0.75\,km\,s$^{-1}$ during post-main-sequence, possibly from asymmetric mass-loss. \citet{Torres2022} have conducted population synthesis fitting of the resolved binary sample within 100\,pc  in order to constrain the binary fraction, initial mass ratio distribution and initial separation distribution. They find the effect of the white dwarf recoil to be weaker than previously proposed by \citet{El-Badry2018}. In addition, they derive a non-flat initial mass ratio distribution of $n(q) \propto q^{-1}$, where $q$ is the mass ratio between the less and more massive components and thus restricted to the range [0,1]. Recent studies have also delved into the dynamic evolution of white dwarfs in triple systems \citep{Shariat2023}. 

The cooling age and mass differences between individual components in wide WD+WD pairs has emerged as a promising metric for testing white dwarf evolution models  \citep{Heintz2022,Hollands2023,Heintz2024}. There is a significant challenge resulting from the steep relation between main-sequence mass and lifetime, which means that canonical $\approx$0.6\,M$_{\odot}$ white dwarfs with long-lived main-sequence progenitors have poorly constrained individual total ages. Hence, white dwarfs with masses $\gtrsim$0.63\,M$_{\odot}$ have been favoured for model age calibration \citep{Heintz2022}.

{\bf Unresolved double degenerates:} \textit{Gaia} has allowed the identification of unresolved double degenerates, primarily focusing on overluminous sources above the white dwarf cooling sequence in the HR diagram. This heightened luminosity arises either from the combined flux of two white dwarfs with similar effective temperatures and masses, or from the primary white dwarf that has lost mass through mass transfer, resulting in an enlarged radius according to the mass-radius relation. Several new candidates for extremely low-mass white dwarfs have been identified prompting subsequent spectroscopic follow-up studies (\citealt{Pelisoli2019};\citealt{Kawka2020}; \citealt{Brown2022}; \citealt{Kosakowski2023}). Double degenerate candidates have also been identified from discrepancies between \textit{Gaia} and spectroscopic mass estimates \citep[see e.g.][]{Tremblay2019,Sahu2023}.

The time-domain and spectroscopic follow-up of overluminous sources in the \textit{Gaia} HR diagram has led to the discovery and characterisation of several short-period, double-lined compact white dwarf systems \citep{Brown2020,Kilic2020_WDWD,Kilic2021,Kilic2021b,Kosakowski2023}, some of which are eclipsing \citep{Keller2022,Munday2023}. These binary systems play a crucial role in testing common-envelope efficiencies \citep{Toonen2012} and predicting the gravitational wave background for space-based gravitational wave detectors such as LISA \citep{LISA1,LISA2}, demonstrating a clear multi-messenger harmony between \textit{Gaia} and the gravitational wave mHz frequency regime \citep{2019MNRAS.490.5888L, Korol2022,Kupfer2023,Ren2023}. In particular, \citet{Korol2022} find a gap at $\approx$1\,au in the double white dwarf separation distribution derived from \textit{Gaia} stellar centroid astrometric wobbling (converted from \textsc{RUWE} parameter), which they attribute to the effect of common-envelope evolution.

{\bf Unresolved WD+MS binaries:} \citet{Rebassa2021} have identified 112 WD+MS candidates within 100\,pc based on their placement in the \textit{Gaia} DR3 HR diagram. However, they note that their selection only represents $\approx$10\% of the total underlying WD+MS population, owing to \textit{Gaia} limitations in detecting WD+MS systems with increasingly cooler white dwarfs or brighter main-sequence stars. Therefore, the exploration of such systems has necessitated additional multi-wavelength UV-\textit{Gaia} HR diagrams \citep{Ren2020,Anguiano2022,Nayak2023}, time-domain eclipse information \citep{Keller2022}, or the detection of astrometric variations \citep{Belokurov2020,Penoyre2022,Shahaf2023a,Shahaf2023,Garbutt2024}. In particular, \citet{Yamaguchi2024} have used \textit{Gaia} DR3 astrometry to characterise a population of WD+MS candidates in relatively wide orbits (100--1000\,d), which likely require a very efficient envelope ejection during common envelope evolution.

{\bf Interacting compact binaries:} The combination of \textit{Gaia}, follow-up spectroscopy \citep[see e.g.][]{Inight2023}, as well as time-domain and X-ray/UV surveys has had a huge impact in identifying and characterising new interacting compact binaries. In particular, \textit{Gaia} astrometry has allowed \citet{Pala2020,Pala2022} to report an updated mass distribution for 89 cataclysmic variable (CV) white dwarfs, finding a mean of $\approx$0.8\,M$_{\odot}$ (compared to $\approx$0.6\,M$_{\odot}$ for field white dwarfs) and no evolution of the mass with orbital period. The \textit{Gaia} science alerts program regularly identifies outbursting CVs \citep{2021A&A...652A..76H}. Additional work has identified a relation between period and position on the HR diagram \citep{Abrahams2022}. \citet{El-Badry2021_CV} presented a survey for recently detached CVs with evolved secondaries, which are proposed to be progenitors of extremely low mass white dwarfs and AM Canum Venaticorum (AM CVn) systems, characterised as hydrogen deficient compact binaries. They find that the implied Galactic birth rate of these CVs is half that of AM CVn binaries. The population of AM CVn binaries has itself been more accurately characterised thanks to \textit{Gaia} astrometry \citep{Ramsay2018,Abril2020,Roestel2022,Kupfer2023}, and selection cuts reliant on \textit{Gaia} have facilitated the discovery of new systems greatly \citep{Roestel2022}, including the sieving of outbursting AM~CVn candidates \citep{2021AJ....162..113V}. 

The varied approaches to identify and characterise compact binaries using \textit{Gaia} data underscore the possibility of creating volume-limited samples encompassing all categories of binaries. These samples aim to be both representative of the broader population and substantial enough to act as a benchmark for future models of binary populations and gravitational wave background \citep{Inight2021,Kawash2022,Canbay2023}.

{\bf White dwarf merger products:} The onset of white dwarf crystallisation in the \textit{Gaia} HR diagram coincides with a distinctive statistical increase in tangential velocities (see Section\,\ref{sec:crystallisation}). \citet{Cheng2019,Cheng2020} used these observations to suggest that $\approx$20\% of massive white dwarfs (0.8-1.3\,M$_{\odot}$) originate from WD+WD mergers, aligning with the predictions from \citet{Temmink2020}. However, recent investigations propose that a portion of these white dwarfs may originate from mergers involving a white dwarf and a subgiant star \citep{Shen2023}. While this particular subset of merger products is predicted to experience an extra crystallisation delay (see Section\,\ref{sec:crystallisation}), the majority of merger products (including WD+WD, WD+MS, MS+MS) are likely to display age and mass distributions close to those of white dwarfs formed from single star evolution \citep{Temmink2020,Kilic2020_100pc}.

Efforts have been made to better characterise the subset of mergers identified by \citet{Cheng2019} through observational follow-up of massive white dwarfs. There is emerging evidence that surface carbon pollution in massive white dwarfs  \citep[hot DQ, warm DQ, DAQ spectral types;][]{Coutu2019,Koester2019,Hollands2020,Kawka2023,Kilic2024} can be explained by convective dredge-up only if the total helium and hydrogen masses are very low \citep{Althaus2009,Hollands2020,Koester2020}, which indicates a merger origin.  Other indicators of mergers include large magnetic fields and rapid rotation \citep{Kilic2023},  although tracing back the evolutionary history of single massive white dwarfs remains challenging.

{\bf Partial thermonuclear explosions:} \textit{Gaia} has played a significant role in characterising remnants arising from incomplete or failed supernova explosions. These remnants emerge when a system, rather than undergoing a complete supernova event, undergoes partial disruption, resulting in the formation of a white dwarf with an exotic O/Ne-dominated surface composition. Additionally, this process may propel both the white dwarf and companion star into a hyper-velocity trajectory that make them unbound from the Galaxy \citep{Shen2018,Raddi2019,El-Badry2023_D6,Igoshev2023,Werner2023}.

We refer the reader to \citet{El-Badry2024rev} for a more extensive review of stellar multiplicity in \textit{Gaia} including compact objects.

\subsection{Magnetic white dwarfs}
\label{sec:magnetic}

\textit{Gaia} has played a transformative role in deriving the parameters of magnetic white dwarfs, constituting approximately 20\% of the local population \citep{Landstreet2019}. This sub-class was historically challenging to study, primarily due to the lack of both parallaxes and a quantum theory framework for combined Stark and Zeeman line broadening. Recent hybrid photometric and spectroscopic studies, relying on \textit{Gaia} parallaxes, have determined robust atmospheric parameters for magnetic white dwarfs \citep[see e.g.][]{Hardy2023,Hardy2023b}.
While fitting magnetic white dwarfs with \textit{Gaia} photometry still faces challenges due to modified line and continuum opacities, the accuracy of \textit{Gaia}-guided fits has been affirmed through UV and IR multi-wavelength HR diagrams. These studies demonstrate that cool magnetic white dwarfs are well-represented by non-magnetic broadband photometric models \citep{Mccleery2020}. Additionally, the observation that both magnetic and non-magnetic white dwarfs crystallise at the same location in the \textit{Gaia} HR diagram suggests not just a mere coincidence but rather similarly accurate \textit{Gaia}-derived stellar parameters for both types.

\begin{figure}
    \centering
    	\includegraphics[width=0.55\textwidth]{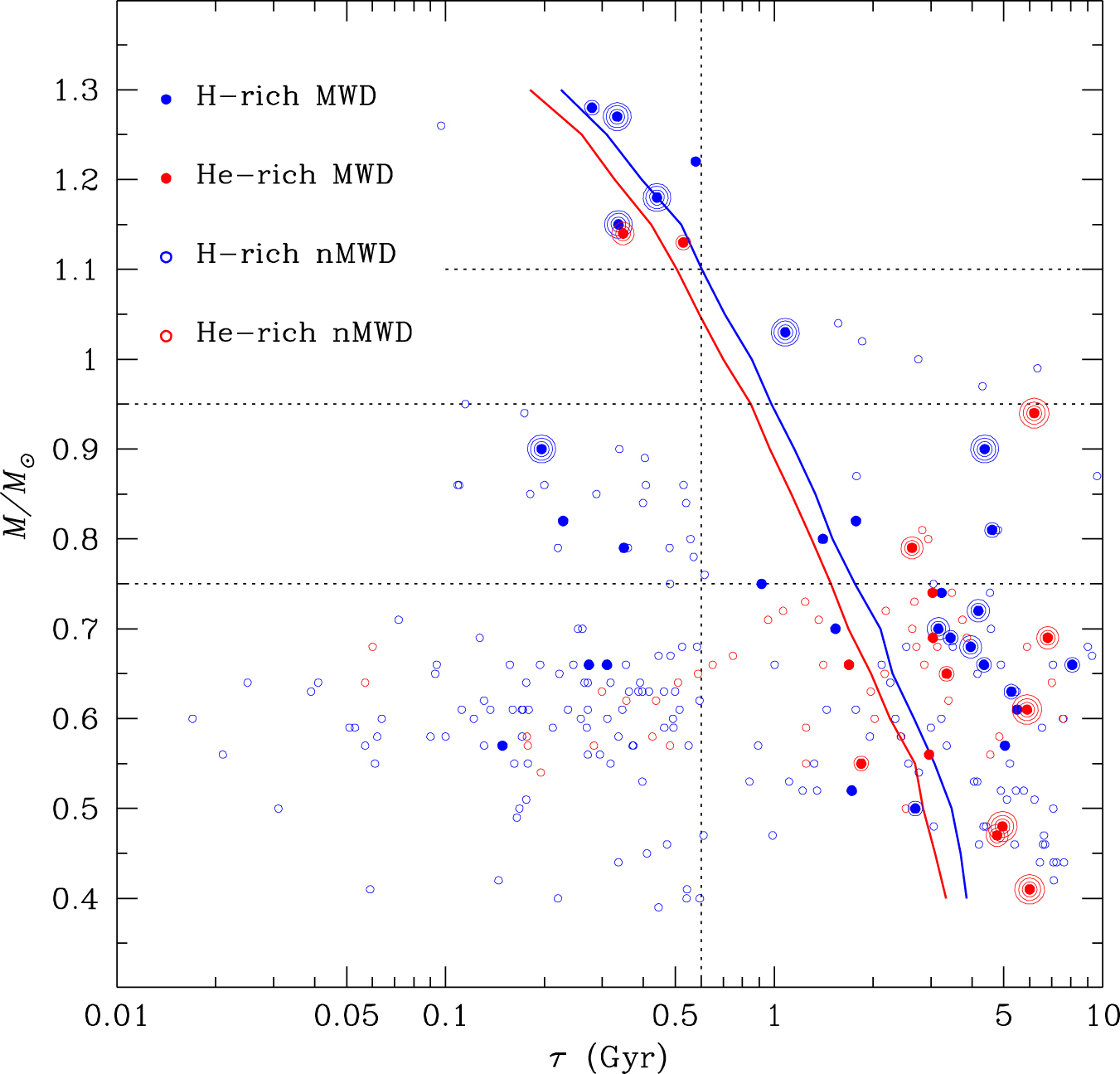}
	\caption{Cooling age-white dwarf mass diagram encompassing all white dwarfs within the local 20\,pc volume and most white dwarfs aged less than 0.6\,Gyr up to a distance of 40\,pc. Unmarked circles represent white dwarfs devoid of detected magnetic fields, while filled dots serve to denote magnetic white dwarfs, where the level of magnetic field strength is indicated by the number of surrounding circles: one for fields below 1\,MG, two for fields between 1 and 10\,MG, and three for fields exceeding 100\,MG. Solid curves trace the onset of core crystallisation for helium-rich (red) and hydrogen-rich (blue) atmospheres (Source: \citealt{Bagnulo2022}).}
    \label{fig:magnetic}
\end{figure}

The newly derived magnetic white dwarf parameters from \textit{Gaia} have validated several previously hypothesised trends, which are discernible in volume-limited samples. Magnetic white dwarfs tend to be more massive than their non-magnetic counterparts \citep{Mccleery2020,Ferrario2020,Bagnulo2021,Hardy2023,O'Brien2024}. Additionally, the incidence of magnetism increases with cooling age for white dwarfs less massive than 0.8\,$M_{\odot}$ \citep[][see Fig.\,\ref{fig:magnetic}]{Bagnulo2022,O'Brien2024}. These behaviours are not fully explained due to the origin of magnetic fields in white dwarfs still being elusive. Possible origins include stellar mergers \citep{Garcia2012}, fossil fields \citep[see e.g.][]{Braithwaite2004}, crystallisation dynamos \citep[see e.g.][]{Isern2017,Ginzburg2022,Montgomery2024,Fuentes2024}, main-sequence and giant-phase internal dynamos with possible external influence (binarity and planets; see e.g. \citealt{Tout2008}; \citealt{Kissin2015}), all of this complicated by a potential delay in the emergence of magnetic fields to the surface \citep{Bagnulo2022}.

A novel spectral class of DA(H)e white dwarfs, which have hydrogen-dominated atmospheres and display Balmer line emission, have been identified through \textit{Gaia} follow-ups \citep{Gaensicke2020,Walters2021,Manser2023,Redding2023,Elms2023,Farihi2023}. This class encapsulates white dwarfs that exhibit strong magnetism ($B>5$\,MG; DAHe), which is found in the majority of cases, and those which lack an observable magnetic field (DAe). DA(H)e white dwarfs lack close brown dwarf or stellar mass companions therefore the prevailing explanation for their observational characteristics points to an intrinsic magnetic phenomenon, potentially linked to the emergence of magnetic fields and the heightened incidence of magnetic white dwarfs for cooling ages exceeding 1--3\,Gyr.

\subsection{Evolved planetary systems}
\label{sec:pollution}

The paradigm attributing the origin of metal atmospheric pollution in white dwarfs to the external accretion of planetary debris was established two decades ago \citep{Jura2003,Zuckerman2007}. \textit{Gaia} data has significantly improved the accuracy of atmospheric chemical abundance determinations for metal-polluted D(A)(B)(Q)Z white dwarfs. This has led to a better understanding of the chemical composition and evolution of planetary systems around evolved white dwarfs \citep{Blouin2019,Coutu2019,Swan2019,Tremblay2020,Klein2021,Izquierdo2021,Hollands2022,Johnson2022,Blouin2022b,OBrien2023,Swan2023,Doyle2023,Rogers2024b,2024MNRAS.527.4515B}.

The spectroscopic follow-up of faint and cool \textit{Gaia} white dwarf candidates has led to the first identification of atmospheric pollution by lithium and potassium \citep{Hollands2021,Kaiser2021,Elms2022,Vennes2024}. Furthermore, the higher than expected concentrations of those elements compared to other atoms has been linked to either the accretion of crust material, the signature of Big Bang nucleosynthesis, or inaccuracies in current atomic diffusion time calculations.

The availability of large \textit{Gaia} white dwarf catalogues has prompted investigations into infrared excesses around white dwarfs resulting from debris disks \citep{Xu2020,Rebassa2019,Lai2021}, gas emission from these discs \citep{Melis2020,Manser2020,GentileFusillo2021_gas}, transiting debris \citep{Guidry2021}, and the frequency of metal pollution in wide binaries \citep{O'Brien2024,Noor2024}. \textit{Gaia} has also stimulated searches for astrometric planet candidates \citep{Andrews2019,Sanderson2022,Kervella2022,Rogers2024}, although this area of research is still in its infancy as raw astrometric data is not yet publicly available. However, it is anticipated to become a significant focus with future data releases.

\section{Astrophysical relations}
\label{sec:astro}

\begin{figure}
    \centering
    	\includegraphics[width=0.6\textwidth]{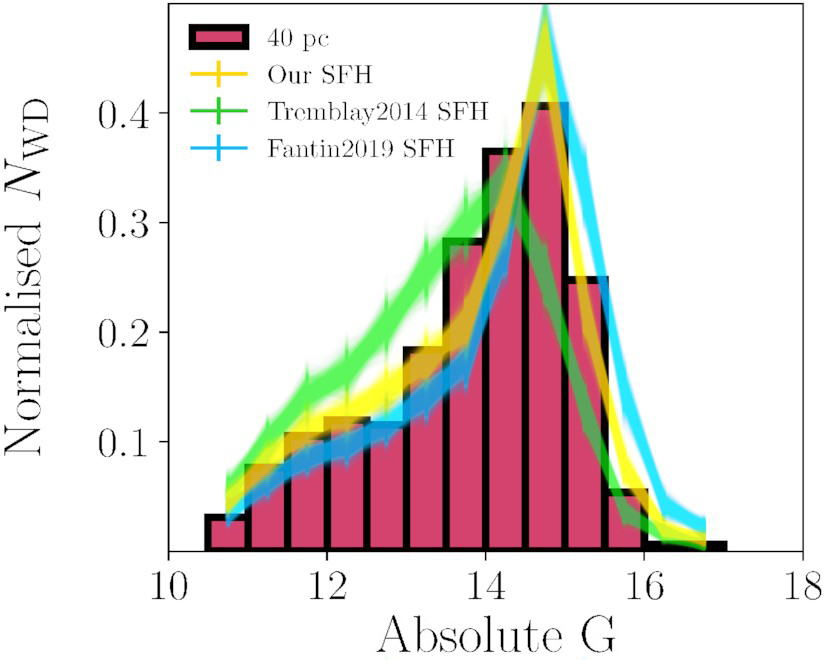}
	\caption{Normalised histogram of the absolute \textit{Gaia} $G$ magnitudes for the 40\,pc volume-limited sample (black). The three contours show the results (and uncertainties) of population synthesis models using three different parameterisation for the star formation rate. A uniform rate for the last 10.6\,Gyr is plotted in yellow. In green and blue are simulated populations relying on the stellar formation history of \citet{Tremblay2014} and \citet{Fantin2019}, respectively (Source: \citealt{Cukanovaite2023}).}
    \label{fig:SFH}
\end{figure}

\begin{figure}
    \centering
    	\includegraphics[width=0.6\textwidth]{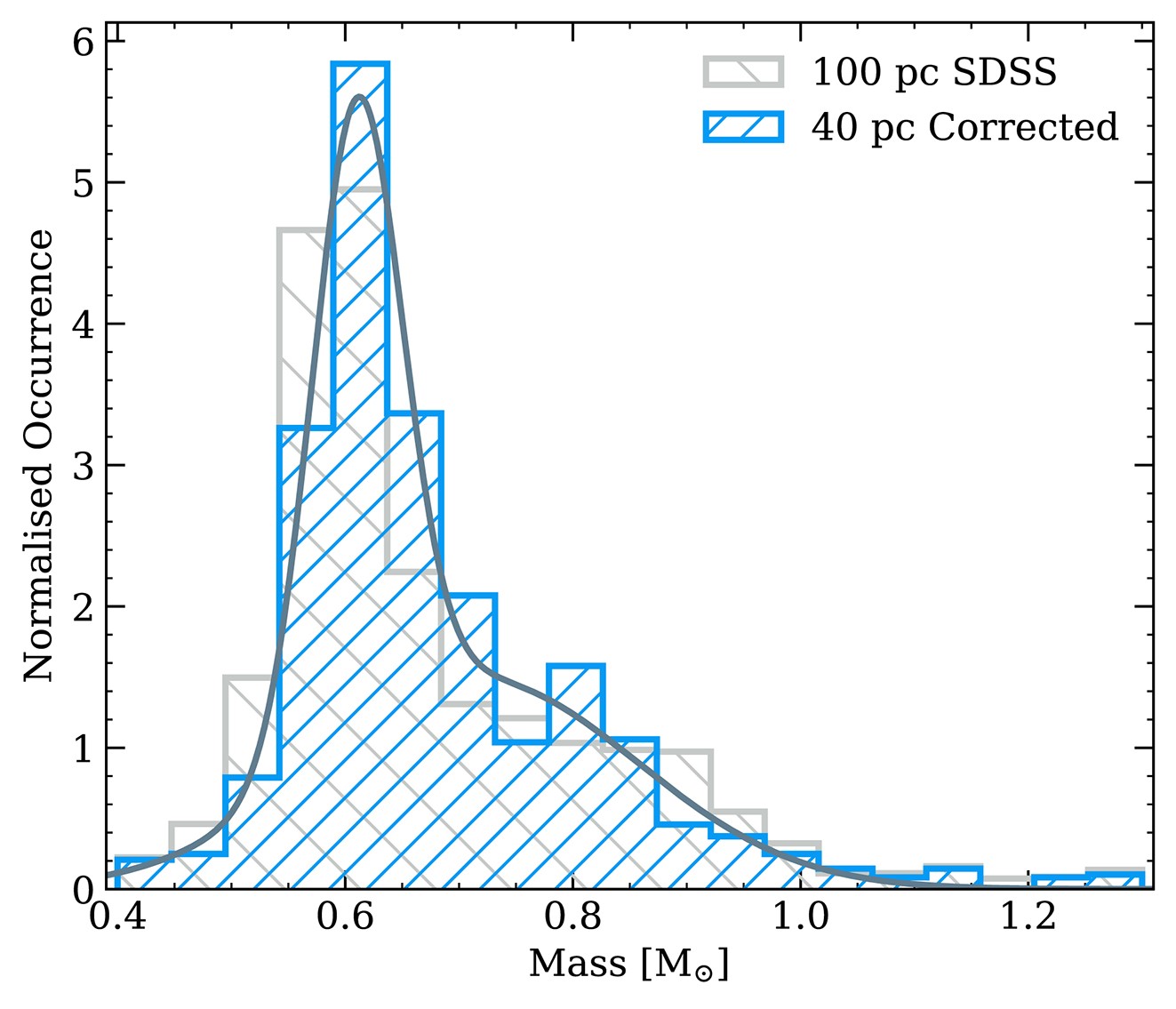}
	\caption{Mass distributions of white dwarfs in the volume complete 40\,pc sample \citep{O'Brien2024} and the 100\,pc SDSS sample \citep{Kilic2020_100pc}. The 40\,pc mass distribution has been corrected for the low-mass problem at low temperatures ($T_{\rm eff} <$ 6000\,K) outlined in Section\,\ref{sec:parameters}. The solid line represents the bimodal best-fitting Gaussians to the 40 pc mass distribution. (Source: \citealt{O'Brien2024}).}
    \label{fig:mass}
\end{figure}

Larger and less biased samples of white dwarfs, along with more precise and accurate stellar parameters, have contributed to the improved calibration of fundamental astrophysical relations using white dwarfs. This process typically begins by modelling the kinematic distribution (Fig.\,\ref{fig:kinematic}), absolute magnitude (or luminosity) distribution (Fig.\,\ref{fig:SFH}) and mass distribution (Fig.\,\ref{fig:mass}) of white dwarfs.

\begin{figure}
    \centering
    	\includegraphics[width=0.8\textwidth]{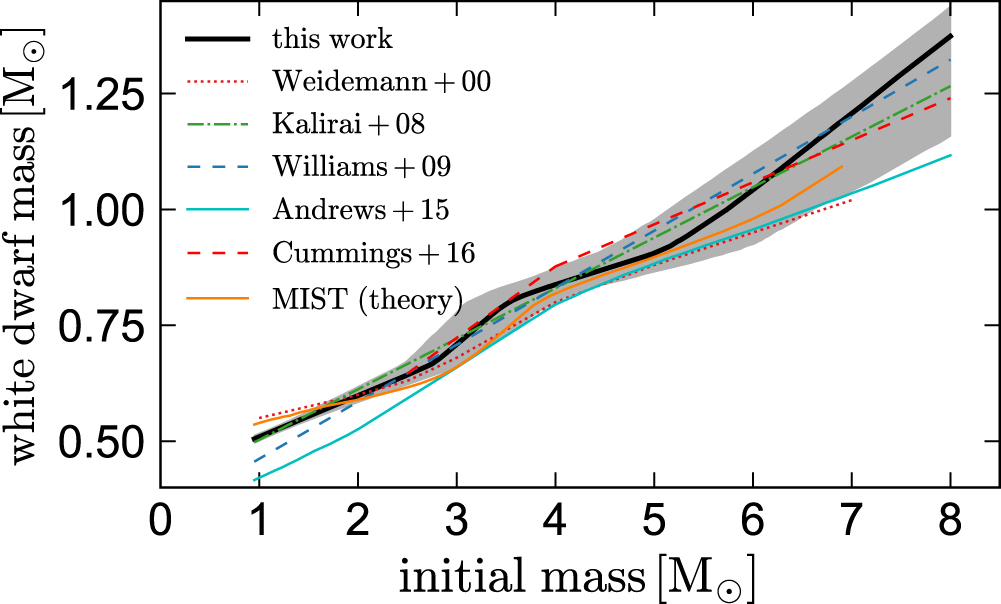}
	\caption{Best-fit population synthesis model initial-final mass relation from the \textit{Gaia} 100\,pc white dwarf sample (``this work" curve and error contour) compared to other results from the literature (Source: \citealt{El-Badry2018}).}
    \label{fig:IFMR}
\end{figure}

The initial-final mass relation (IFMR), which connects the masses of the progenitor main-sequence star and the white dwarf, is pivotal for constraining mass loss in stellar evolution and determining the total age of white dwarf populations. Leveraging the \textit{Gaia} white dwarf mass distribution in volume-limited samples (Fig.\,\ref{fig:mass}), \citet{El-Badry2018} and \citet{Cunningham2024} have introduced a novel statistical method to derive the IFMR (Fig.\,\ref{fig:IFMR}). This technique ensures that the predicted total age of a white dwarf (time since zero-age main sequence) is self-consistent with white dwarf masses determined from \textit{Gaia} data and aligns with our general understanding of Galactic disc age, star formation rate and initial mass function (IMF).

Equally important studies have employed \textit{Gaia}-characterised wide binaries \citep{Barrientos2021,Hollands2023} and stellar clusters \citep{Marigo2020} to obtain more accurate IFMRs. The advantage of these latter two techniques lies in their ability to characterise the scatter in the IFMR by incorporating external constraints on the total age of each individual white dwarf, such as the age of the cluster or wide companion. Although the (possibly intrinsic) scatter in the IFMR persists at the $\approx$5\% level \citep{Hollands2023}, the median IFMR is statistically equivalent across most studies and white dwarf masses (see Fig.\,\ref{fig:IFMR}).

The \textit{Gaia} HR diagram has been used to deduce the local star formation rate in the Galactic disc, employing both stars and white dwarfs. In the latter case, \citet{Cukanovaite2023} utilised the \textit{Gaia} absolute magnitude distribution of 40\,pc white dwarfs, demonstrating its consistency with a constant star formation rate over the last $\approx$10.6\,Gyr (see Fig.\,\ref{fig:SFH}). However, alternative studies have instead suggested star formation peaks at either early or late times \citep{Fantin2019,Isern2019,Fleury2022}. Explorations of star formation histories have also expanded to include the Galactic halo, demonstrating that white dwarfs serve as a suitable tool for inferring halo age and local space density \citep{Kilic2019,Torres2021,Fantin2021}, although halo age remains sensitive to systematic uncertainties in stellar evolution models. In particular, \citet{Kilic2019} and \citet{Torres2021} derive a Galactic inner halo age of 10.9 $\pm$ 0.4\,Gyr and 12 $\pm$ 0.5\,Gyr, respectively.

Coupling the local star formation history and the IFMR is the IMF. While it has historically been studied using young stellar clusters that have not yet formed white dwarfs, recent studies have constrained the IMF using the local and globular cluster populations including white dwarfs \citep{Dickson2023,Kirkpatrick2024}, finding consistency in the 1--8\,$M_{\odot}$ initial-mass range with the long-established Kroupa and Salpeter IMFs.

White dwarf kinematics have been investigated using \textit{Gaia} proper motions and external radial velocity measurements. This has been employed to derive the age versus velocity dispersion relation and Galactic radial migration history \citep{Torres2019,Rowell2019,Mikkola2022,Raddi2022,Cukanovaite2023}. 
These relations are independent and complementary to those derived from local main-sequence stars, and are critical to extract the local star formation history \citep{Cukanovaite2023}. Due to their motions through space, the large majority of \textit{Gaia} white dwarfs currently within 100\,pc of the Sun have formed further away. Understanding the kinematic origin of the local white dwarf sample allows us to place its stellar formation history into the context of Milky Way evolution. In a recent study, \citet{Zubiaur2024} use \textit{Gaia} Galactic orbital parameters to find that 68\% of current 100\,pc white dwarfs were stars born within 1\,kpc from the Sun.

Given that white dwarfs lack original metallicity information, precise radial velocities play a crucial role in making white dwarfs a tracer of Milky Way evolution \citep{Raddi2022}. Directly measured astrometric \textit{Gaia} radial velocities is a promising avenue for extracting 3D velocities of white dwarfs with weak spectral signatures \citep{Lindegren2021}. \textit{Gaia} white dwarfs have also proven useful in age determinations of coeval and comoving young local stellar associations \citep{Torres2019,Gagne2020}.

Assuming that white dwarf ages are better, or at least as well understood as the ages of local main-sequence stars, the former have been employed to calibrate several main-sequence age relations, including isochrone fitting, age-velocity dispersion, age-activity and age-metallicity relations \citep{Fouesneau2019,Qiu2021,Moss2022,Heintz2022,Rebassa-Manserga2021b,Zhang2023,Rebassa-Mansergas2023}. Despite these efforts, there remains a significant scatter between different methods for determining main-sequence and white dwarf ages, most prominently seen in the WD+MS and WD+WD wide binary samples \citep{Qiu2021,Heintz2024}.

In open clusters, \textit{Gaia} has facilitated the identification of new white dwarfs and confirmation of existing members \citep[see e.g.][]{Griggio2022}. This allowed refining age determinations of well known nearby open clusters such as the Hyades and Praesepe \citep{Salaris2018,Salaris2019,Lodieu2019b,Lodieu2019}, identifying new magnetic or exotic white dwarfs in clusters, suggesting that magnetic white dwarfs may form in isolation \citep{Caiazzo2020}, and providing new insights into the high-mass end of the IFMR \citep{Cluster2021,Richer2021,Heyl2022,OpenCluster,Griggio2022,Miller2023}. \textit{Gaia} has also been useful in improving HR diagrams of globular clusters including white dwarfs \citep{Sahu2022}.

\textit{Gaia} white dwarf catalogues have also enhanced our understanding of local stellar extinction, particularly in determining GALEX UV reddening coefficients \citep{Wall2019}. Accurate extinction measurements are crucial for both white dwarf selection \citep{Gentile2021} and obtaining accurate atmospheric parameters \citep{Sahu2023}.

Finally, \textit{Gaia} data have played a crucial role in testing the white dwarf mass-radius relation through the combination of spectroscopic, photometric and gravitational redshift measurements \citep{Joyce2018,Tremblay2019,Genest-Beaulieu2019,Arseneau2023,Pasquini2023}. The mass-radius relation has also been constrained via a combination of \textit{Gaia} and \textit{HST} astrometric microlensing \citep{McGill2018,McGill2023}. These studies have confirmed that current theoretical mass-radius relations are appropriate \citep{Renedo2010,Romero2019,Bedard2020,Salaris2022}.

\section{Future}
\label{sec:conclusion}

The number of high-probability \textit{Gaia} white dwarf candidates has experienced a 38\% increase between DR2 and DR3, and this trend is expected to continue with upcoming data releases. Despite the identification of over 355\,000 white dwarf candidates in \textit{Gaia} DR3, only approximately 10\% have undergone medium-resolution spectroscopic follow-up \citep{Gentile2021}. As new multi-object spectroscopic surveys such as DESI, WEAVE, SDSS-V and 4MOST observe more white dwarfs in the coming decade, the scientific yield from existing \textit{Gaia} data releases is anticipated to further increase, particularly for the rare and exotic spectral classes discussed in this review.

The combination of \textit{Gaia} epoch photometry and astrometry in future data releases with external time-domain surveys like LSST \citep{Fantin2020} and ZTF \citep{ZTF} will enable the identification of more variable white dwarf systems. This includes compact binaries as sources of gravitational waves, giant planet companions down to one Saturn mass at orbital separations of 0.1--50\,au \citep{Kervella2022,Rogers2024}, and transiting rocky planetary debris \citep{Guidry2021}.
\\
\\

\noindent{\bf Acknowledgement:} This work was supported by the European Research Council under the European Union’s Horizon 2020 research and innovation programme numbers 101002408 a Leverhulme Trust Research Grant
(ID RPG-2020-366).

\bibliographystyle{model2-names-astronomy} 
\bibliography{main}

\end{document}